\documentstyle[epsfig,12pt]{article}
\font\twelve=cmbx10 at 15pt
\font\ten=cmbx10 at 12pt

\renewcommand{\thefootnote}{\fnsymbol{footnote}}

\begin{document}

\def\Z{\hbox{ Z\hskip -9pt Z}}
\def\R{\hbox{\it I\hskip -3pt R}}
\def\N{\hbox{\it I\hskip -3pt N}}

\begin{titlepage}

\begin{center}

{\ten Centre de Physique Th\'eorique\footnote{Unit\'e Propre de
Recherche 7061} - CNRS - Luminy, Case 907}

{\ten F-13288 Marseille Cedex 9 - France }

\vspace{2cm}

{\twelve KINKS DYNAMICS IN ONE-DIMENSIONAL COUPLED MAP LATTICES}

\vspace{0.3cm}
\setcounter{footnote}{0}
\renewcommand{\thefootnote}{\arabic{footnote}}
{\bf Bastien FERNANDEZ}

\vspace{3cm}

{\bf Abstract}

\end{center}

We examine the problem of the dynamics of interfaces in a
one-dimensional space-time discrete dynamical system. Two different
regimes are  studied: the non-propagating and the propagating one. In
the first case, after  proving the existence of such solutions, we
show how they can be described using Taylor expansions. The second
situation deals with the assumption of a travelling wave to follow the
kink propagation. Then a comparison with the corresponding  continuous
model is proposed. We find that these methods are useful in simple
dynamical situations but their application to complex dynamical
behaviour is not  yet understood.

\vspace{3 cm}

\noindent Number of figures : 3

\bigskip

\noindent October 1994

\noindent CPT-94/P.3079

\bigskip

\noindent anonymous ftp or gopher: cpt.univ-mrs.fr

\end{titlepage}

\section{Introduction}
Coupled Map Lattices (CML) have been proposed as simplest models of
extended dynamical systems and they serve now as a paradigm in the
analysis of complex  nonlinear phenomena. CML is a discrete space,
discrete time, continuous state  dynamical system that is
deterministic \cite{Kaneko84} \cite{Kaneko85a}
\cite{Crutchfield87}. These models are easy to handle numerically but
the  mathematical analysis is not so simple. Also they often display
the essential features of physical systems (for a review, see for
example \cite{chaos}). The numerical studies show that CML give rise
to a rich phenomenology including  a wide variety of both spatial and
temporal periodic structures, intermittency,  chaos,  domain walls,
kink dynamics, {\sl etc.} \cite{Kaneko93} \cite{Farmer}. Some of
these phenomena were  also investigated from a rigourous mathematical
point of   view \cite{Bunimovich88} \cite{Bunimovich92}
\cite{Amritkar}.
\smallskip

The problem that we are dealing with is the study of particular
localized  coherent structures in CML, the interfaces between two
different phases, the  so-called kinks or fronts. Depending on the
nature of the phases, these  interfaces can
emerge in a simple or rather complex dynamical context. In the former
case, kinks  appear when both (stationary) phases are in competition
as is the case in  alloy solidification \cite{Kessler} or in the
propagation of chemical waves
\cite{Richard}. Interfaces in such systems can be either propagating
or  non-propagating fronts depending on the stability of both phases
and on the value
 of the control parameter, here taken as the diffusion coefficient.
The  dynamical behaviour is said to be simple because of the low
temporal period and  the (quasi-) homogeneity of the different phases,
namely the domains. On the  other hand, in spatiotemporal
intermittency, phases are more complex solutions (in  the sense of
periodicity) and they have been called the natural wavelengths. Kinks
 in this case play again a central role as their motions destabilize
these  wavelengths and thus lead to the well-known bursts of
turbulence inherent in
 intermittency \cite {Farmer} \cite{Lambert92}.
\smallskip

The goal of this article is to give a rigourous description of the
dynamics of  the kinks that may occur in CML models. The advantage of
CML is based  on the use of local maps with well-controled dynamical
properties. Tuning  the parameters in such systems enables the
observation and study of various regimes of the front dynamics.
Moreover it gives a discrete alternative  point of view to the more
conventional tool of partial differential equations (PDE)  in which
the dynamics of localized structures, at least for some PDE, begins
to  be well understood \cite{Collet90} \cite{Kawasaki}.
\smallskip

We shall compare the behaviours of these interfaces in the discrete
space-time  dynamical system and the corresponding continuous one.
After introducing the  model and patterns under consideration, we
concentrate on the situation of  non-propagating solutions in Section
3. We show the existence and unicity (up to translations in the
lattice) of such kinks in the small coupling range. Its  expression is
then evaluated using Taylor expansions. In a second step, the
propagating fronts are analyzed with the assumption of being a
travelling wave  as was already done in PDE \cite{Schlogl80}. In this
case there are important  differences between discretized and
continuous models concerning the dynamics of
 interfaces and we address this issue in Section 4. Our assumptions
are tested  numerically. Finally an extension of the preceding results
to upper temporal
 period is presented and we end with some concluding remarks.

\section{Definitions}
In this section we begin the description  the model using the general
definition given in \cite{Bunimovich93}.
\smallskip

Let $\Z\ $ be the (infinite) unidimensional lattice. At each site $i
\in \Z\ $ the local  phase space is $X_i$ (here $X_i$ is chosen as
either the interval $[0,1]$ or
$[-1,1]$). The phase space of our dynamical system is the direct
product
$M={\displaystyle \prod_{i \in \Z}} X_i$. Consequently a point $x \in
M$ can be  written $x=(x_i)_{i \in \Z}$.
\smallskip

CML is a mapping $\Phi_{\epsilon}$ of $M$ into itself depending on
the coupling  parameter $\epsilon$. As we are dealing with a model of
reaction-diffusion systems, $\Phi_{\epsilon}$ splits into the
composition of two mappings
$$\Phi_{\epsilon}=G_{\epsilon}\circ F \eqno (2.1)$$

\noindent
where $(Fx)_i=f(x_i)$ is the {\sl reaction}, that is to say the
action of a local  nonlinear mapping $f:X_i\longrightarrow X_i$ and
$(G_{\epsilon}x)_i=g_{\epsilon, i}(x)$ is the interaction. In our case
this interaction is chosen to be  representative of the {\sl
diffusion} in the simplest way which reads
$$g_{\epsilon,i}(x)=x_i+{\epsilon \over 2}(x_{i+1}+x_{i-1}-2x_i)$$

\noindent
where the coupling parameter $\epsilon \in [0,1]$. Finally the new
state of the CML at time $t+1$ is given by the following convex linear
combination (of the state  at time $t$):
$$x_i^{t+1}=(\Phi_{\epsilon} x^t)_i=(1-\epsilon)f(x_i^t)+{\epsilon
\over 2}
\Bigl( f(x_{i+1}^t)+f(x_{i-1}^t) \Bigr) \eqno (2.2)$$

\noindent
which implies the location of $x^{t+1}$ in M as soon as $x^t \in M$.
\medskip

\noindent
Let $\{X^1,X^2,\cdots,X^p\}$ denote either $p$ different fixed points
or a  period-$p$ orbit of $f$.
\smallskip

\noindent
{\bf Definition:} A left domain (respectively a right domain) is a
sequence
$x=\{x_i\}_{i=-\infty}^{i=N_1}$ ({\sl resp.} $x=\{x_i\}_{i=N_2}^{i
=+\infty}$)  such that there is an $X^p$ such that

\noindent
$|x_i^t-X^p|\leq \alpha$ for all $i \leq N_1$ ({\sl resp.}
$|x_i^t-X^p|\leq
\alpha$ for all $i \geq N_2$)

\noindent
where $\alpha$ is a small positive parameter to be fixed later.
\medskip

\noindent
The different values $X^p$ are referred to as phases of the domain. A
kink is the
 interface between a left and a right domain with different phases
({\sl fig.} 1).  A rigorous definition of these patterns can be given
with $\alpha=0$ \cite{Bunimovich92}, but this implies the use of
local  maps that are constant over an open subset of $X_i$ and this
constant is a point
 of the considered periodic orbit. In our case, we deal with more
 general maps that do not have these peculiar features. Thus,
 it is not possible to find kink-like solutions with domain values
(strictly) equal to one of the
$X^n$. In other words domains do not (really) exist when the lattice
reveals the presence of a  kink and when $\epsilon$ is non-zero. This
explains the above definition where
$\alpha$ can be thought of as a
control parameter for separating the interface from the coherent
domains. The  smaller is $\alpha$, the longer is the interface.

\section{Non-propagating kinks}
We address in this section the problem of non-propagating interfaces.
After proving the existence of such kinks in the  small coupling
range, we show how the use of Taylor expansions allows us to describe
these solutions and induces
 the reduction of the dynamics to a finite dimensional space.
\smallskip

For the sake of simplicity, the map $f$ is assumed to be $k$-times
continuously differentiable with bounded  derivatives and also to have
two stable fixed points $X_1$ and $X_2$. Thus the domains under
consideration are fixed  structures for the CML dynamics and the kinks
are inspected as fixed points.
\smallskip

We are looking for the solutions of $\Phi_{\epsilon}(x)=x$ with the
boundary conditions ${\displaystyle \lim_{i\rightarrow -\infty}} x_i
=  X_1$ and ${\displaystyle \lim_{i\rightarrow \infty}} x_i = X_2$.
When $\epsilon=0$ there exists a trivial solution $x^*$ and it is easy
to see  that $x^*$ is invariant under translations on the lattice.
This solution is  written
$$x^*=(\cdots,X^1,X^1,X^2,X^2,\cdots)$$

\noindent
Now we endow the phase space $M$ with the sup norm: $\| x\| =
{\displaystyle
\sup_{i \in \Z}} |x_i|$. One can check that $(M,\|.\|)$ is a closed
ball of a  Banach space\footnote{We will keep the notation $M$ for
this Banach space.}. Moreover the Fr\'echet derivative of
$\Phi_0$ at $x^*$, $D\Phi_0(x^*)$, is a bijective operator from $M$
into $\R^{\Z}$ since it is diagonal and none of its  elements
vanishes. Finally the operators $\Phi_{\epsilon}$ and $D\Phi_0$ are
bounded and continuous at $x^*$.

\noindent
Thus by the Implicit Function Theorem, we obtain the existence and
unicity of the  kink-like fixed point in the neighborhood of $\epsilon
= 0$, that is to say in a  certain interval of strength of space
interactions $[0,\epsilon'[$.  We may note that, in the more general
case of temporally periodic domains of period $p$, this theorem can
also be invoked using $(\Phi_{\epsilon})^p$ instead of
$\Phi_{\epsilon}$ to prove the existence and unicity of the kink-like
periodic-p orbit.
\bigskip

Since we know the existence of the interface when $\epsilon$ is
small, we can now describe it. A first step in this problem is to
follow the kink genesis  from a perturbative point of view. The system
is initialized in the following  state:
$$\left\{\matrix{
\forall i \leq s&x_i^0=X^1\cr
\forall i > s   &x_i^0=X^2\cr} \right. \eqno(3.1)$$

\noindent
where $s$ is fixed in $\Z$ . (3.1) is the so-called step.
\smallskip

\noindent
This pattern is submitted to the dynamics and the successive states
are computed  using expansions (close to $X^1$ if $i\leq s$, to $X^2$
otherwise) in terms of
$\epsilon$ powers. When these expansions are limited to the $k^{th}$
power,  it can be proved that $x^t\ (t>k)$ is given by
the system:
$$\left\{\matrix{
x_i^t&=X^1 \hfill
                         &           &\hfill i\leq&-t+s\hfill\cr
x_i^t&=X^1+o(\epsilon^k) \hfill
                         &\hfill -t+s&<i\leq\hfill&-k+s\hfill\cr
x_i^t&=X^1+{\displaystyle \sum_{l=s-i+1}^k}a_{i,l}(t)\epsilon^l+o(
\epsilon^k) \hfill&\hfill -k+s&<i\leq\hfill&s   \hfill\cr
x_i^t&=X^2+{\displaystyle \sum_{l=i-s}^k}a_{i,l}(t)\epsilon^l+o(
\epsilon^k) \hfill  &\hfill    s&<i\leq\hfill&k+s \hfill\cr
x_i^t&=X^2+o(\epsilon^k)\hfill  &\hfill k+s&<i\leq\hfill&t+s \hfill\cr
x_i^t&=X^2 \hfill  &\hfill  t+s&<i    \hfill&          \cr} \right.
\eqno(3.2)$$

\noindent
where $o(\epsilon^k) $ stands for the remainders of the expansion. In
this  expression, $a_{i,l}(t)$ is the $l^{th}$ power coefficient for
the polynomial  determining $x_i^t$. This general system manifests
itself for any local map and  the diffusive features only enter in the
coefficients. It reveals a kink where  both domains are defined by
$\alpha = o(\epsilon^k)$.
\smallskip

\noindent
Such a solution can be understood in the following way: in the small
coupling  range the perturbative effects (from the local map point of
view) induced by  the presence of the interface are rapidly damped in
space. This explains the  limited number of polynomials appearing in
the characterization of the  structure.
\smallskip

\noindent
{}From the system (3.2) we deduce the interface length expressed in
terms of the  number of intersites: $L_d(k)= 2k$. In order to display
this kink length as a  function of the coupling parameter, one has to
use the relation $\alpha \simeq
\epsilon^k$. We obtain obviously
$$L_d(\epsilon) \simeq {{2 Log \alpha} \over {Log \epsilon}}$$

\noindent
We describe here a first comparison with continuous systems. From a
PDE model  the interface length reads \cite{Kawasaki}
$$L_c(\epsilon) \simeq \sqrt{\epsilon}$$

\noindent
Thus both lengths vanish when $\epsilon$ goes to zero.  However their
comparison tells  us that $L_c$ decreases more rapidly than $L_d$.
There then appears a difference  between the models, starting without
diffusion and increasing slowly the  coupling parameter, according to
both definitions of an interface, we get a  longer kink in a
continuous model than in a discrete one.
\smallskip

\noindent
Furthermore the structure of (3.2) is invariant under the CML
dynamics, that is  to say only the $\epsilon$ polynomial coefficients
evolve in time. In fact, for any  given $k$ and $s$, substituing the
corresponding (3.2) expression in (2.2) using the
 Taylor expansion (until the $k^{th}$ power) for $f$ proves the
invariance.
\bigskip

The kink pattern is therefore given by a set of polynomials with
time-dependent  coefficients. So when studying such structures, the
CML dynamics can thus be  replaced by a map defined on these
coefficients. This substitution allows the  reduction from an infinite
number of relations (in the latter dynamics) to just a  few (in the
coefficient map). In other words the stability problem is now located
in a finite dimensional space which is an issue to the closure problem
in such  unbounded media.
\smallskip

\noindent
The coefficient map analysis is made here for $s=0$ and $k=2$ since it
is the  simplest case and since the other cases can be treated in the
same way. When the
 expansions are limited to the second power, there are six
non-vanishing  coefficients for which the dynamic $F$ is (defined in
$\R^6$)
$$\left\{\matrix{
a_{-1,2}(t+1)&=&\bigl(a_{-1,2}(t)+{a_{0,1}(t) \over 2}\bigr) f'(X^1)
\hfill                                        \cr
a_{0,1}(t+1) &=&X+a_{0,1}(t)f'(X^1) \hfill                       \cr
a_{0,2}(t+1) &=&\bigl(a_{0,2}(t)-a_{0,1}(t)\bigr)f'(X^1)+a_{1,1}(t)
{f'(X^2) \over 2}+a_{0,1}^2(t){f''(X^1) \over 2}\cr
a_{1,2}(t+1) &=&\bigl(a_{1,2}(t)-a_{1,1}(t)\bigr)f'(X^2)+a_{0,1}(t)
{f'(X^1) \over 2}+a_{1,1}^2(t){f''(X^2) \over 2}\cr
a_{1,1}(t+1) &=&-X+a_{1,1}(t)f'(X^2) \hfill                      \cr
a_{2,2}(t+1) &=&\bigl(a_{2,2}(t)+{a_{1,1}(t) \over 2}\bigr) f'(X^2)
\hfill                                       \cr} \right. \eqno(3.3)$$

\noindent
where $X={\displaystyle {X^2-X^1 \over 2}}$ and $f',f''$ denote
successives  derivatives with respect to $x$. Expression (3.3) is
deduced from the  introduction of the system (3.2) in the CML
dynamics. The iterations for
$a_{-1,2}$ and $a_{0,1}$ (respectively $a_{2,2}$ and $a_{1,1}$) are
in this  particular case the relations linking the coefficients.
\smallskip

Before looking at the map $F$, we make a comment on the influence of
$f$  orbits on the kink dynamics. If, instead of choosing a local map
with two fixed  points, we work with a period-2 orbit $\{ X^1,X^2\}$,
then the interface  would flip at every iteration as the domains'
values alternate. The kink structure  would therefore be given by the
system (3.2) for even times and by a  symmetrical (with respect to the
index $s$) expression for odd times. Such a  solution describes a
period-2 orbit for the lattice dynamics. But as the  coefficients'
expressions in this case are quite complicated we refrain from
presenting them here and we will just mention the final results.
\smallskip

\noindent
Notice that these expressions are $s$-independent because of the
translational  invariance in the lattice frame.
\smallskip

The analysis of the map $F$ shows the existence of a fixed point
called
${\cal A}^\infty$ for which the coordinates are
$$\left\{\matrix{
a_{-1,2}^\infty&=&{\displaystyle {Xf'(X^1) \over 2(1-f'(X^1))^2}}
\hfill \cr
a_{0,1}^\infty &=&{\displaystyle {X \over 1-f'(X^1)}}   \hfill \cr
a_{0,2}^\infty &=&{\displaystyle {-X \over 1-f'(X^1)}\biggl( {f'(X^1)
\over 1-f'(X^1)}+{f'(X^2)\over 2(1-f'(X^2))}-{Xf''(X^1)\over
2(1-f'(X^1))^2}\biggr) }\cr
a_{1,2}^\infty &=&{\displaystyle {X \over 1-f'(X^2)}\biggl( {f'(X^2)
\over 1-f'(X^2)}+{f'(X^1)\over 2(1-f'(X^1))}+{Xf''(X^2)\over
2(1-f'(X^2))^2}\biggr) }\cr
a_{1,1}^\infty &=&{\displaystyle {-X \over 1-f'(X^2)}} \hfill \cr
a_{2,2}^\infty &=&{\displaystyle {-Xf'(X^2) \over 2(1-f'(X^2))^2}}
\hfill \cr} \right. $$
\smallskip

\noindent
One can find ${\cal A}^\infty$ when looking for a solution of
$A=F(A),\ A\in
\R^6$.
\smallskip

\noindent
We now compute the jacobian matrix associated to ${\cal A}^\infty$ as
its  spectrum determines the stability of ${\cal A}^\infty$. When
solving the
${\cal A}^\infty$-spectral problem we found three eigenvalues equal
to $f'(X^1)$ andthree others equal to  $f'(X^2)$. Since $X^1$ and
$X^2$ are two stable fixed points, ${\cal A}^\infty$ is stable.
Moreover the map $F$ is strictly contracting so  this fixed point is
unique. Thus the expression (3.2) defines the unique  attractor of the
CML dynamics for the considered expansions.
\smallskip

\noindent
To conclude, the kink stability is only due to local map stability;
 once the domains are installed, there is a unique stable pattern
possible for  the kink as much as small diffusive interactions are
taken into account. This is
 important because it proves that no special preparation is needed to
create and
 stabilize such interfaces; they arise in a natural way, no matter
wath the  initial condition. One can see that this technique not only
confirms the results  given by the Implicit Function Theorem used at
the begining of this section, but also gives an explicit expression
for the patterns under consideration. The  validity domain of these
solutions will be made more precise  later on with a localization
argument.
\smallskip

\noindent
A similar calculus, pursued in the case $k=3$, generates a
coefficient dynamical  system in $\R^{12}$ and leads to the same
conclusion. This technique can also be  used to examine the case of
greater periodicity as said before. One has to  compute expansions in
different phases of the dynamics. Other coefficient maps  are found.
We summarize here some of the results. When $f$ has a periodic orbit
of period two (for $k=2$ and $k=3$) the jacobian matrix associated to
the CML periodic orbit is shown to have only one eigenvalue, namely
$f'(X^1).f'(X^2)$.  In the case $k=2$ this eigenvalue has multiplicity
6 and 12 when $k=3$. The map  is also contracting which allows in
these cases the same conclusions as above.  An extension of this
result to any finite temporal period needs more calculus and is still
in progress.

\section{Propagating fronts}
We now consider a particular aspect of the kink dynamics, namely its
displacement along the lattice. As already stated, this phenomenom
appears as one of the keys of the spatio-temporal intermittency in
CML. As in the fixed interface  situation we begin by investigating
the case where $f$ has two stable fixed points,  then we generalize to
greater temporal periods.
\smallskip

A first step in this direction is made using computer simulations.
They suggest that these interfaces are moving without being deformed
and with a constant velocity  when the local behaviour is quite
simple. Starting from this statement, we assume
 the existence of such a kink-like travelling wave in CML. If that is
the case,  the lattice state at the date $t$ is given by
$$x_i^t=h(i-vt)\ \ \forall t \in \N \ \forall i \in \Z\eqno(4.1)$$

\noindent
where we suppose $h \in C^1(\R,I)$ and $v$ is the travelling
velocity. For the  present purpose, assumptions for the local map are
kept identical with those made  in Section 3. The continuity implies
the existence of a third fixed point
$X_u\ (X^1<X_u<X^2)$ which is unstable \cite{Collet80}. Likewise we
impose the  boundary conditions for $h$ in the co-moving frame
$z=i-vt$:
$${\displaystyle \lim_{z \rightarrow -\infty}} h(z)=X^1, \ \ \
{\displaystyle \lim_{z \rightarrow +\infty}} h(z)=X^2$$

\noindent
We call $z_u$ the point such that: $X_u=h(z_u)$.
\medskip

If a slow motion (as seen in simulations) is considered, that is if
$v \ll 1$,  in one time step the site values $x_i$ stay close to each
other. Consequently we  write in the first order approximation
$$h(z-v)=h(z)-v {d h \over d z}\bigg|_{z=i-vt} \eqno(4.2)$$
\smallskip

\noindent
In a second step the two right-hand sides of the relations (2.2) and
(4.2) are  compared. The expression obtained by generalization to all
real values of the  variable $z$ is
$$h(z)-vh'(z)=f\bigl[ h(z) \bigr]+{\displaystyle \epsilon \over 2}
\Delta f\bigl[ h(z) \bigr] \eqno(4.3)$$

\noindent
Eqn. (4.3) is now intergrated over $\R$. From the $h$ hypothesis we
deduce
$${\displaystyle \int_{-\infty}^{+\infty}} h'(z)\ dz=X^2-X^1$$
\smallskip

\noindent
and: $${\displaystyle \int_{-\infty}^{+\infty}} \Delta f[h(z)]\ dz=0$$
\smallskip

\noindent
Finally the moving speed is given by:
$$v={\displaystyle {1 \over {X^2-X^1}} \int_{-\infty}^{+\infty}}\
\bigl(h(z) - f\bigl[ h(z) \bigr]\bigr) \ dz \eqno(4.4)$$
\smallskip

\noindent
This relation indicates that the kink displacement is a consequence
of a lack of symmetry in the interface shape
 and its image under $f$. We also notice from (4.4) that a necessary
condition  for $v$ to vanish is that the functions $f$ and $h$ are
antisymmetric functions.
 A similar calculus  performed in a continuous space-time model of
chemical  reactions had lead to similar conclusions \cite{Schlogl80}.
\smallskip

Following these comments, this inhomogeneous solution is now studied
for a  peculiar local map, say, the cubic map $f(x)=\nu\ x\ (1-x^2) \
(1<\nu<2)$. As  this reaction is antisymmetric, the kink is expected
also to be antisymmetric and
 thus to be a stationary solution. In order to compute an analytic
expression for this pattern,  we use a continuous description of our
model assuming kinks in CML can be  described by the following
equation (See {\sl e.g.} \cite{Albano84})
$$h(z)=f(h(z))+{\epsilon \over 2}{d^2 h(z) \over dz^2} \eqno(4.5)$$

\noindent
One can find the solutions of (4.5) using an analogy with the
equation of  motion of a particle of {\sl mass} ${\epsilon \over 2}$
at {\sl position} $h$,  in {\sl time} $z$ evolving in the potential
$$V(h)=\int_I (f(h)-h)dh={\nu-1 \over 2} h^2-{\nu \over 4}h^4$$

\noindent
Investigating a first integral of motion (the constant is given by
the above  boundary conditions) leads to the kink profile in the
stationary case
$$h(z)=\sqrt{\nu -1 \over \nu} \tanh \biggl( \sqrt{\nu-1 \over
\epsilon} z \biggr) \eqno(4.6)$$

\noindent
A plot of this function verifies the validity of the approximation
(4.5) ({\sl fig.} 2). One can obtain a better fit with the discrete
kink when considering instead of  (4.5) the equation
$$h(z)=f(h(z))+{\epsilon \over 2}{d^2 f(h(z)) \over dz^2}$$

\noindent
The mechanical analogy is also applicable in this model but no simple
analytic  expression for $h$ was found.
\smallskip

\noindent
In any case we take (4.6) as a starting point for understanding the
non-symmetric  case and the single map is chosen to be the {\sl
perturbed} cubic map
$f(x)=\nu\ x\ (1-x^2) + p\ \ (p \ll 1)$. We suppose that the effect
of the  perturbation is to induce a displacement of the interface. So
we make again the ansatz of  the travelling wave (which has been
proved for a certain class of partial  differential equations
\cite{Saarloos}):
$$h(i,t)=C\ p+\sqrt{\nu -1 \over \nu} \tanh \biggl( \sqrt{\nu-1 \over
\epsilon} (i-p.v.t) \biggr) \ \ (C \in \R) \eqno(4.7)$$

\noindent
Eqn. (4.7) is substituted in (4.5) also supposing a small speed. For
the first order term this yields
$$v= -{3 \over 2} {\sqrt{\epsilon \nu} \over \nu -1} \eqno (4.8)$$

\noindent
It is also possible to compute the constant $C$ but, since it plays a
secondary  role in this solution, we do not give its expression here.
More interesting is the
 comparison among the expressions (4.4) and (4.8) and the measured
speed from  numerical simulations ({\sl fig.} 3). The gaps between the
experimental data for the  velocity are due to the finite size of the
lattice used in the numerical  experiment. Notice that there is an
acceptable fit for large $\epsilon$, while we
 can see a difference between the continuous model and the CML
behaviour when the
 diffusion plays a small role. This is because the approximation made
in (4.5)  is no longer valid when $\epsilon$ decreases, as discrete
effects become more and  more important. In other words, the discrete
laplacian operator cannot be  considered as a diffusion operator in
the classical sense in the small coupling  range as its variations
along the front are much more important and, in a certain sense, less
regular than those of the corresponding continuous operator.
Furthermore this figure also shows that the assumption (4.1)
concerning the  existence of a travelling wave in CML is in good
agreement with the experimental  results for any strength of space
interactions. This is confirmed by a plot of the  velocity
distribution when the kink propagates in space and by the expansion
method in the fixed case. Indeed, it is proved that computing the
velocity $v$  using (4.4) and the interface shape given by system
(3.2) with asymptotic  coefficients always gives a vanishing value.
However the proof of the existence  of such a travelling wave solution
in CML is still an open problem.
\smallskip

We may understand in a more detailed way the difference between CML
and a  continuous model. One has to display a discrete argument to
explain a non-moving  interface when the strength of space
interactions is small. With the above  assumptions on the single map,
there exists an attracting neighbourhood near  each fixed point such
that $f$ restricted to these open subsets is stricly  contracting. We
shall see that the interface is stationary if each $x_i^t$  remains
located in one of the intervals. The following points are defined:
$X_{Inf}$ (resp. $X_{Sup}$) such that $f'(X_{Inf})=1$ and
$X^1<X_{Inf}<X_u$  (resp. $f'(X_{Sup})=1$ and $X_u<X_{Sup}<X^2$). Now
we consider
$O_1 = ]X^1,X_{Inf}[$ and $O_2 = ]X_{Sup},X^2[$ as these open
attracting  subsets:
\medskip

\noindent
{\bf Proposition:} There exists an $\epsilon_c$ such that for all
$\epsilon <
\epsilon_c$ we have for all $t \geq t_0$
$$\left\{\matrix{
\forall i \leq N&x_i^t \in O_1\cr
\forall i > N   &x_i^t \in O_2\cr} \right. $$

\noindent
if this condition is true at $t=t_0$ and if $f$ is an increasing
function on
$]X^1,X^2[$. $N$ depends on the initial condition.
\medskip

\noindent
The proof is direct by induction. It is enough to write the different
inequalities at time $t$ and to submit them to the dynamics using
relation (2.2)
 and the properties: $f(X_{Inf})<X_{Inf}$ and $f(X_{Sup})>X_{Sup}$. A
necessary  condition for the location in $O_1$ (respectively in $O_2$)
gives the value:
$\epsilon_1={X_{Inf} - f(X_{Inf})\over X^2 - f(X_{Inf})}$ (resp.
$\epsilon_2={f(X_{Sup})-X_{Sup} \over f(X_{Sup})-X^1}$). Finally we
get
$\epsilon_c=\min \lbrace \epsilon_1,\epsilon_2 \rbrace$.
\smallskip

\noindent
In the case of the perturbed cubic map we compute the value
$\epsilon_c \simeq  0.13$ which is in good agreement with the
experimental data. From this proposition  we also obtain an effective
range of application of the expansion method. This value $\epsilon_c$
may also serve as an upper  bound for the $\epsilon'$ given by the
Implicit Function Theorem. This bound is  up to now the only mark in
the study of the transition from a non-propagating to   a propagating
interface in our model (with a continuous nonlinear map).
\smallskip

In a more general dynamical context, that is to say when $f$ has a
period-$p$  orbit, the analysis can be done in a similar way. In this
situation we  assume the existence of a periodic sequence of
travelling waves $\{  h_1(z),\cdots,h_p(z) \} $ describing the CML
states, each $h_i(z)$ being again  continuously differentiable. The
generalized boundary conditions are
$$\left\{\matrix{
h_n^{+\infty} \equiv {\displaystyle\lim_{i \rightarrow +\infty}}
h_n(i) = X^{n+1} \hfill \cr
h_n^{-\infty} \equiv {\displaystyle\lim_{i
\rightarrow -\infty}} h_n(i) =     X^n     \hfill \cr} \right. $$

\noindent
with the periodic condition: $X^{p+1}=X^1$.

\noindent
Following the technique employed in the simplest case and assuming
that the speed  is independent of the phase of the solution then
yields
$$v={\displaystyle {1 \over {h_n^{+\infty}-h_n^{-\infty}}}
\int_{-\infty}^{+\infty}}\ \bigl(h_n(z) - f\bigl[ h_n(z) \bigr]\bigr)
\ dz \eqno(4.9)$$
\smallskip

\noindent
We have tried to test this expression in cases where the period-2 and
-4 orbits  of the logistic map are stable. In both cases the system
reveals a non-moving  interface for any value of the diffusive
coupling. This unexpected observation  was confirmed by the velocities
computed from (4.9). This computed velocity  vanishes whatever the
phase $n$ we consider in (4.9).
\smallskip

We end the study of propagating kinks with a remark on the most
complex  behaviour, that is to say, an interface between natural
wavelengths. In order to
 obtain such a situation, one has to choose a single map with an
attracting set  having two disjoint intervals exchanging their
dynamics \cite{Farmer}
\cite{Lambert92}. Then the preceding reasoning can be pursued with
slightly  different boundary conditions that are periodic in time, and
it gives an  expression very similar to (4.9). However a comparison
with numerical experiments is  quite difficult in this realistic
situation. We have noted from simulations that  the spatial
oscillations induce shape modifications of the moving structures, the
wavelengths destabilize themselves and irregular behaviour is observed
during a
 transient before the stabilization of a (different) wavelength. Thus
all the  assumptions required of $h$ are no longer valid. For
instance, it is  easy to see that the shape is deformed under
iterations and the  differentiability is not respected. We found
however some examples of  agreement between theoretical and numerical
results. They are not given here  since they do not constitute an
effective confirmation of the hypothesis.

\section{Summary and discussion}
In this paper we have considered the problem of a peculiar type of
inhomogeneous  solution in CML. First the use of a technique based on
expansions to describe  stationary interfaces has revealed itself as
an efficient tool and we concluded  that it yields to a stable
solution in all case. The second goal was to show how a  travelling
wave could properly describe propagating kinks. In order to obtain  an
analytic expression for its shape, a continuous description has proved
its  usefulness even if it is no longer valid in the small coupling
parameter range.  At first sight one can conclude that there is a
contradiction between both results,  but a localization argument was
used to link these apparently conflicting assertions.
\smallskip

\noindent
We have demonstrated the efficiency of two different tools in the
study of
 the kink dynamics in CML. When compared to the case of the PDE, we
have seen that the effect of the discretization is to shift the
critical
 value for the transition between stationary and travelling kinks,
{}from
$\epsilon=0$ in PDE to $\epsilon_c >0$ in CML. In other words there
appears a  vanishing speed in the small coupling range even if $f$ has
no peculiar symmetry.
 Our method was shown to generalize to the case of period-$p$ orbits.
Finally  in spatio-temporal intermittency, the validity of such a
description is still an
 open question.
\bigskip

\noindent
{\bf Acknowledgements}

I thank P. Collet and S. Ruffo for their comments during the {\sl
Workshop on Dynamics and Complexity} (Lisboa, 09/1994). I also thank
R. Lima, A.  Lambert and E. Ugalde for fruitfull discussions.

\vfill\eject


\begin{thebibliography}{99}




\bibitem{Kaneko84} K. Kaneko. Period-doubling of kink-antikink
patterns, quasi-periodicity in antiferrolike structures in coupled
logistic lattice. {\it Prog. of Theo. Phys.}, 72:480-486, 1984.

\bibitem{Kaneko85a} K. Kaneko. Spatial period doubling in open flow.
{\it Phys. Lett. A}, 111:321-325, 1985.

\bibitem{Crutchfield87} J.P. Crutchfield and K. Kaneko.
Phenomenology of spatio-temporal chaos. In Directions in Chaos,
editor, {\it Hao Bai-Lin}, pages 272-353. World Scientific,
Singapore, 1987.

\bibitem{chaos} Chaos focus issue on coupled map lattices. Chaos
{\bf 2}, 1992.

\bibitem{Kaneko93} K. Kaneko. The coupled map lattices. In Theory
and Applications of Coupled Map Lattices, editors, {\it K. Kaneko}.
Wiley, 1993.

\bibitem{Farmer} J.D. Keeler and J.D. Farmer. Robust space-time
intermittency and ${1 \over f}$ noise. {\it Physica D}, 23:413-435,
1986.

\bibitem{Bunimovich88} L.A. Bunimovich and YA.G. Sinai. Spacetime
chaos in coupled map lattices. {\it Nonlinearity}, 1:491-516, 1988.

\bibitem{Bunimovich92} L.A. Bunimovich, R. Livi, G. Martinez-Mekler,
and S. Ruffo. Coupled trivial maps. {\it Chaos}, 2(3):283-291, 1992.

\bibitem{Amritkar} R.E. Amritkar and P.M. Gade. Spatially periodic
orbits in coupled map lattices. {\it Phys. Rev. E}, 47(1):143, 1993.

\bibitem{Kessler} D.A. Kessler, J. Koplik, and H. Levine. {\it Adv.
Phys.}, 37:255, 1988.

\bibitem{Richard} Oscillations and Travelling Waves in Chemical
Systems, editors. {\it R.J. Field and M. Burger}. Wiley, New York,
1985.

\bibitem{Lambert92} A. Lambert and R. Lima. Stability of wavelenghts
and spatio-temporal intermittency. {\it Physica D}, 71(4):390, 1994.

\bibitem{Collet90} P. Collet and J.P. Eckmann. {\it Instabilities
and Fronts in Extended Systems}. Princeton University Press, 1990.

\bibitem{Kawasaki} K. Kawasaki and T. Ohta. Kink dynamics in
one-dimensional nonlinear system. {\it Physica A}, 116:573-593, 1982.

\bibitem{Schlogl80} F. Schl\"ogl and R. Stephen Berry. Small
roughness fluctuations in the layer between two phases. {\it Phys.
Rev. A}, 21(6):2078-2081, 1980.

\bibitem{Bunimovich93} L.A. Bunimovich. Coupled map lattices: One
step forward and two steps back. preprint, 1993.

\bibitem{Collet80} P. Collet and J.P. Eckmann.
{\it Iterated Maps on the interval as a Dynamical System}.
Birkhauser, Boston, 1980.

\bibitem{Albano84} A.M. Albano and N.B. Abraham and D.E.
Chyba and M. Martelli. Bifurcations, propagating solutions and phase
transitions in a nonlinear chemical reaction with diffusion. {\it
Am. J. Phys.}, 52(2):161-167, 1984.

\bibitem{Saarloos} W.V. Saarloos. Front propagation into unstable
states: Marginal stability as a dynamical mechanism for velocity
selection. {\it Phys. Rev. A}, 37(1):211-229, 1988.

\end{thebibliography}

\vfill\eject
\noindent
{\Large{\bf Figures Captions}}
\medskip

\noindent
{\bf Figure 1}: Snapshot of a CML of 30 sites (horizontal axis) after
10000  iterations with $\epsilon = 0.25$. The local map is the
logistic map $f(x)=\mu
\ x\ (1-x)$ with $\mu=3.2$. The dashed lines represent domains
values, namely
$X_1$ and $X_2$. The lattice has been initialized with the so-called
step  (see text).
\medskip

\noindent
{\bf Figure 2}: Snapshot of a CML of 25 sites (horizontal axis) after
10000  iterations of the cubic map, $\nu = 1.4$ and $\epsilon = 0.6$.
The dashed line  is the corresponding {\sl continuous} solution (4.6)
\medskip

\noindent
{\bf Figure 3}: Plot of the travelling velocity $v$ {\sl v.s.} the
coupling
$\epsilon$. The local map is the perturbed cubic map with $\nu = 1.3$
and $p  =0.02$. Squares stand for the measured velocity obtained with
a lattice of 100  sites, the solid line is the one computed
(numerically) by the integration  formula (4.4) while the dashed line
represents theorical velocity obtained in  the continuous model, Eq
(4.8). The negatives values stand for a front  propagating toward the
left.

\vfill\eject

\begin{figure}
\epsfxsize=12cm
\centerline{\epsffile[50 50 770 554]{vis_lattp.ps}}
\end{figure}

\begin{figure}
\epsfxsize=12cm
\centerline{\epsffile[50 50 770 554]{check_port.ps}}
\end{figure}

\begin{figure}
\epsfxsize=12cm
\centerline{\epsffile[50 50 770 554]{vitesse.ps}}
\end{figure}

\end{document}